\documentclass[final,5p,times,twocolumn]{elsarticle}

\usepackage{graphicx}
\usepackage{graphics}
\usepackage{amssymb}

\begin{document}

\begin{frontmatter}

\title{The Lovelock gravity in the critical spacetime dimension}

\author[nkd1,nkd2]{Naresh Dadhich}
\ead{nkd@iucaa.ernet.in}

\author[nkd1]{Sushant G. Ghosh}
\ead{sgghosh2@jmi.ac.in}

\author[nkd1]{Sanjay Jhingan\corref{cor1}}
\ead{sanjay.jhingan@gmail.com}

\address[nkd1]{Centre for Theoretical Physics, Jamia Millia Islamia,
New Delhi 110025, India}

\address[nkd2]{Inter-University Centre for
Astronomy \& Astrophysics, Post Bag 4, Pune 411 007, India}

\cortext[cor1]{Sanjay Jhingan}

\begin{abstract}
It is well known that the vacuum in the Einstein gravity, which is
linear in the Riemann curvature, is trivial in the critical
$(2+1=3)$ dimension because vacuum solution is flat. It turns out that this is true in general for any
odd critical $d=2n+1$ dimension where $n$ is the degree of
homogeneous polynomial in Riemann defining its higher order
analogue whose trace is the $n$th order Lovelock
polynomial. This is the "curvature" for $n$th order pure Lovelock
gravity as the trace of its Bianchi derivative gives the
corresponding analogue of the Einstein tensor \cite{bianchi}. Thus
the vacuum in the pure Lovelock gravity is always trivial in the odd
critical $(2n+1)$ dimension which means it is pure Lovelock flat but
it is not Riemann flat unless $n=1$ and then it describes a field of
a global monopole. Further by adding $\Lambda$ we obtain the
Lovelock analogue of the BTZ black hole.
\end{abstract}

\end{frontmatter}


Since gravity is universal as it links to everything that physically
exists including zero mass particles, hence it can only be described
by the curvature of spacetime and its dynamics is then entirely
determined by the Riemann curvature. The Einstein-Hilbert action
which is the trace of the Riemann curvature tensor gives on
variation the second rank symmetric Einstein tensor with vanishing
divergence. The Einstein tensor provides the second order
differential operator, the analogue of $\nabla^2\phi$, in the
equation of motion. There is however an alternative purely geometric
way to get to the Einstein tensor by taking the trace of the Bianchi
identity satisfied by the Riemann curvature. Inclusion of higher
order terms in curvature in the action becomes pertinent to take
into account the high energy effects. That is we have to go beyond
the linear Einstein-Hilbert term to a polynomial in Riemann and the
requirement of the second order quasi-linear equation uniquely
identifies the polynomial to the Lovelock polynomial. On the
alternative geometric side we have to find an analogue of the
Riemann tensor which is a polynomial in Riemann. The Riemann
satisfies the Bianchi identity which means vanishing of its Bianchi
derivative and the trace of the identity leads to the divergence
free Einstein tensor. Now the higher order analogue of Riemann as
identified by Dadhich in \cite{bianchi} has non-zero Bianchi
derivative and hence does not satisfy the Bianchi identity which is
the defining property of Riemann tensor. However the trace of the
Bianchi derivative does indeed vanish and that is what is required
to get to the divergence free analogue of the Einstein tensor. The
trace of the higher order Riemann analogue is indeed the Lovelock
polynomial but for the numerical multiplying factor.

It is well known that for the linear in Riemann Einstein gravity,
vacuum is trivial in $3$ spacetime dimension as $R_{ab}=0$ implies
$R_{abcd}=0$. There exists no non-trivial vacuum solution to
incorporate dynamics. The vacuum solution is non-trivial only in
dimension $\geq4$. To universalize this feature, we should ask
whether it is true in general for higher order gravity as well? That
is, is vacuum solution trivial in general for the critical $d=2n+1$
dimension relative to the higher order Riemann analogue where $n$ is
the degree of the polynomial? If we denote $n$th order Riemann
analogue by $R^{(n)}_{abcd}$, then $R^{(n)}_{ab}=0$ implies
$R^{(n)}_{abcd}=0$ for $d=2n+1$ and the $n$th order pure Lovelock
vacuum will be non-trivial only in $d\geq2(n+1)$ dimension. Even
when spacetime is Lovelock flat, it  will not be Riemann flat unless
$n=1$.

Our main purpose in this paper is to establish this universal
feature of gravitational field. This means spacetime dimension and
the degree of the curvature polynomial, $R^{(n)}_{abcd}$, are
intimately related and $d=2n+1$ is the critical dimension for which
the corresponding vacuum is trivial. For the linear and quadratic
orders $n=1,2$, it is the Einstein and Gauss-Bonnet gravity with
critical dimensions $d=3,5$ respectively. And vacuum is universally
trivial in the critical dimensions. The Lovelock flat is not Riemann
flat unless $n=1,$ and the static spacetime in the critical
dimension is characterized by $g_{tt}=-1/g_{rr}=const.$. Then
$g_{tt}$ could be squared out by redefining the time coordinate as
constant Newtonian potential is trivial while $g_{rr}$ is the
Einstein effect which cannot be absorbed by coordinate
transformation and it represents a solid angle deficit for $d>3$
and has non-zero Riemann curvature \cite{dadhich1}. The Einstein
stress tensor so generated is known asymptotically to approximate to
that of a global monopole in $4$ dimension \cite{bv,dny}. It is
remarkable that this is true in general for all dimensions $
> 3$.
We shall in particular show that the Gauss-Bonnet trivial vacuum
spacetime in the critical $5$ dimension indeed produces Einstein
stresses that describe a $5$-dimensional global monopole. Thus
Lovelock flat spacetime in the critical dimension $(2n+1)$ will
describe a global monopole in the Einstein gravity.

Following Dadhich \cite{bianchi} with a slight change in notation we
define the Lovelock curvature polynomial
\begin{eqnarray}\nonumber
R^{(n)}_{abcd} &=& F^{(n)}_{abcd}  - \frac{n-1}{n(d-1)(d-2)} F^{(n)} (g_{ac}g_{bd} - g_{ad}g_{bc}), \\
F^{(n)}_{abcd} &=& Q_{ab}{}{}^{mn} R_{cdmn}
\end{eqnarray}
where
\begin{eqnarray}\nonumber
Q^{ab}{}{}_{cd} &=& \delta^{a b a_1 b_1...a_n b_n}_{cdc_1 d_1...c_n
d_n} R_{a_1 b_1}{}{}^{c_1 d_1}...R_{a_{n-1} b_{n-1}}{}{}^{c_{n-1} d_{n-1}}, \\
Q^{abcd}{}{}{}{}_{;d}&=&0.
\end{eqnarray}
The analogue of $n^{th}$ order Einstein tensor is given by
\begin{equation}
G^{(n)}_{ab} = n(R^{(n)}_{ab} - \frac{1}{2} R^{(n)} g_{ab})
\end{equation}
 and
\begin{equation}
 R^{(n)} = \frac{d-2n}{n(d-2)}F^{(n)}
\end{equation}
Note that $R^{(n)}=R^{(n)}_{ab}g^{ab}=0$ in $2n$ dimension for
arbitrary metric $g_{ab}$. Since $R^{(n)}_{ab}$ is a function of the
metric and its first and second derivatives which are all arbitrary,
it must vanish in $d=2n$. That is, $R^{(n)}_{ab}=0$ identically in
$2n$ dimension. On the other hand for the general Lovelock case, the
lagrangian is non-zero for $d=2n$ but its variation vanishes
identically. Here it is much more direct and transparent . Further
it turns out that
\begin{equation}
R^{(n)}_{abcd} =\Lambda (g_{ac} g_{bd} - g_{ad} g_{bc})
\end{equation}
implies
\begin{equation}
F^{(n)}_{abcd} = \frac{n(d-2)}{d-2n}\Lambda (g_{ac} g_{bd} - g_{ad} g_{bc})\;,
\end{equation}
and vice versa. Not only that the corresponding Weyl curvature is also the same for the two.
That is
\begin{eqnarray}
W^{(n)}_{abcd} (R^{(n)}_{abcd}) &=& R^{(n)}_{abcd} -\frac{1}{(d-2)} \nonumber\\
&\times &  \left(
R^{(n)}_{ac} g_{bd} + R^{(n)}_{bd} g_{ac} - R^{(n)}_{ad} g_{bc} - R^{(n)}_{bc} g_{ad}\right) \nonumber \\
&+& \frac{1}{(d-1)(d-2)}R^{(n)} (g_{ac} g_{bd} - g_{ad} g_{bc}) \nonumber \\
&=& W^{(n)}_{abcd}(F^{(n)}_{abcd}) \;.
\end{eqnarray}
The two tensors differ from each other only through their trace.

We shall now explicitly demonstrate for the static spacetime that
pure Lovelock vacuum,  $G^{(n)}_{ab}=0$, solution is in fact
Lovelock flat, $R^{(n)}_{abcd}=0$. We write for the static
spherically symmetric spacetime,
\begin{equation}
ds^2 = B dt^2 - A dr^2 - r^2 d\Omega_{(d-2)}^2
\end{equation}
where $AB = const. =1$ due to the null energy condition, $G^{(n)}_{ab}k^a k^b=0, k_a k^a=0$. Then the vacuum solution with
$\Lambda$, $G^{(n)}_{ab}=\Lambda g_{ab}$, is given by \cite{vac},
\begin{equation}\label{solA}
B = 1/A = 1 - \left(\Lambda r^{2n} + \frac{M}{r^{d-2n-1}}\right)^{1/n}.
\end{equation}
In the critical dimension $d=2n+1$, the pure Lovelock vacuum
solution with $\Lambda=0$  will have $B=1/A=1-K=const.$ which could
however be transformed away in $g_{tt}$ but not in $g_{rr}$ for
$d > 3$ and hence is Riemann non flat. However it will have
$R^{(n)}_{abcd}=0$ for any $n$. It is trivially true for $n=1$
because it only causes the angle deficit which produces no Riemann
curvature and we have verified it for the Gauss-Bonnet case, $n=2$.
This shows that the pure Gauss-Bonnet vacuum is trivial; i.e.
$G^{(n)}_{ab}=0$ implies $R^{(n)}_{abcd}=0$ in the critical
dimension, $d=2n+1$ which is $5$ in this case. Similarly it could be
verified for any $n$ \footnote{Since the appearance of the
first version on the arXiv, there appeared a paper \cite{kastor} the very
next day establishing this result in general for any $n$. Thus vacuum for
the critical dimension $d=2n+1$ is always Lovelock flat.}. It is remarkable that in critical dimension
spacetime is characterized by the vanishing of the corresponding
$n^{th}$ order curvature. Thus for the critical $d=2n+1$ dimension,
pure Lovelock vacuum is always Lovelock flat but it would not be
Riemann flat unless $n=1$. We now show that it would describe a
global monopole in the Einstein gravity.

Since the Lovelock flat spacetime in the critical dimension is not
Riemann flat, hence it will  generate the Einstein stresses in the
equation,
\begin{equation}
G_{ab} = - \kappa T_{ab}.
\end{equation}
For the critical $5$-dimensional Gauss-Bonnet vacuum we have $B= 1/A
= 1-K = const.$ which  gives rise to Einstein stresses,
\begin{equation}
G^t_t = G^r_r = 3G^\theta_\theta = -3\frac{K}{r^2}, \quad G^{\theta}_{\theta} = G^{\phi}_{\phi} = G^{\psi}_{\psi}.
\end{equation}
To felicitate comparison with the four dimensional global monopole
solution we have used the same notation as that of Barriola and
Vilenkin \cite{bv} and write the Lagrangian as,
\begin{equation}\label{lag}
{\cal L} = \frac{1}{2} \partial_{\mu} \phi^a \partial^{\mu} \phi^a-\frac{1}{4}\lambda (\phi^a \phi^a - \eta^2)^2 \;,
\end{equation}
where $\phi^a$ is a quadruplet of scalar fields $(a=1,2,3,4)$. The
field configuration  describing monopole is
\begin{equation}\label{field}
\phi^a = \eta f(r)\frac{x^a}{r}
\end{equation}
where $x^a$ are cartesian coordinates with the usual relation to
spherical  coordinates, and $x^a x^a = r^2$. The energy momentum
tensor of monopole then takes the form
\begin{eqnarray}\label{emtensor}\nonumber
T^t_t &=& \frac{1}{2A^2} \eta^2 {f'}^2 +\frac{3}{2}\frac{\eta^2
f^2}{r^2} + \frac{\lambda}{4}\eta^4(f^2-1)^2 \\
T^r_r &=& -\frac{1}{2A^2} \eta^2 {f'}^2 +\frac{3}{2}\frac{\eta^2
f^2}{r^2} + \frac{\lambda}{4}\eta^4(f^2-1)^2 \\ \nonumber
T^{\theta}_{\theta}
&=&\frac{1}{2A^2} \eta^2 {f'}^2 +\frac{1}{2}\frac{\eta^2
f^2}{r^2} + \frac{\lambda}{4}\eta^4(f^2-1)^2 \;
\end{eqnarray}
and due to spherical symmetry $T^{\theta}_{\theta}=T^{\phi}_{\phi} =
T^{\psi}_{\psi}$.  The equation of motion for the field $\phi^a  $
reduces to the following equation for $f(r),$
\begin{equation}\label{feq}
\frac{f''}{A} +\left[\frac{3}{r A}+\frac{1}{2B}
\left(\frac{B}{A}\right)'\right] f' -\frac{3f}{r^2} -\lambda\eta^2f
(f^2-1)=0 \; .
\end{equation}

As is clear from the above discussion, the critical dimension vacuum
spacetime can harbor  no core mass for global monopole, and
asymptotically $f\approx1$ and then we have
\begin{equation}
T^t_t = T^r_r = 3 T^{\theta}_{\theta} = \frac{3\eta^2}{2 r^2}.
\end{equation}
We can now integrate the Einstein equation to write the metric
coefficient in the exterior  as given by
\begin{equation}
B= 1/A = 1 - \frac{8 \pi G_5}{r} \int{T^t_t r^2 dr} = 1 - 12 \pi G_5 \eta^2.
\end{equation}
The size of the monopole core could be estimated in flat space as
$\delta \approx \lambda^{-1/2} \eta^{-1}$. This approximation is
believed to hold good as gravity is not expected to substantially
alter the structure of the monopole. Our aim is to show
that the stresses generated by $B = 1/A = const.$ has the same
structure as that of the global monopole and the constant $K=12\pi
G_{5} \eta^2$.

We had set out to establish the two remarkable universal features:
(a) the universality of vacuum in the critical $d=2n+1$ dimension in
which spacetime is free of the corresponding curvature
$R^{(n)}_{abcd}$; i.e. "vacuum is flat" and (b) this spacetime
always describes a global monopole in the Einstein gravity. That's
what we have shown. The critical dimension vacuum spacetime could be
viewed as due to constant Newtonian potential which geometrically
corresponds to solid angle deficit. The remarkable point is that it
produces stress structure which agrees with that of a global
monopole not only in $4$ dimension \cite{bv} but also in any
dimension $d\geq4$. This is very interesting, why should the
stresses always match? It is though understandable that the stresses
go as $1/r^2$ because that is what the solid angle deficit could do
and so does the prescription of the field $\phi^a$. However what is
not so obvious is the fact that in $4$ dimension all the angular
stresses vanish but not in $5$ dimension yet the stresses exactly
match for the left and right of the equation. What is interesting
here is the fact that a global monopole in the critical dimension
$(2n+1)$ in the Einstein gravity is in fact a trivial vacuum
solution relative to the Lovelock gravity with vanishing
corresponding curvature, $R^{(n)}_{abcd}$. Alternatively we can also
view it as a constant potential spacetime which is Lovelock flat in
the critical dimension.

If we do not set $\Lambda=0$ in the solution (\ref{solA}), it would
describe the analogue of BTZ black hole \cite{btz} in the critical
$d=2n+1$ dimension. Note that the BTZ black hole is the solution of
$G^{(n)}_{ab} = \Lambda g_{ab}$ in the critical dimension, $d=2n+1$
and hence it exists only in the critical dimension. Thus BTZ black
hole with all its peculiar and remarkable properties exists in all
critical dimensions with the corresponding "curvature"
$R^{(n)}_{abcd}$ being constant. Though BTZ black hole is well known
but what is not so well known is the property that it occurs not
only in $3$ dimension but in all odd critical dimensions and its
spacetime is indeed of constant curvature, $R^{(n)}_{abcd}$. This we
believe is a new feature that has got uncovered through our higher
order curvature analysis.

The main motivation for this investigation was to explore the
universal features of gravity in higher dimensions. Starting from
the universality of gravity inside uniform density sphere \cite{dmk}
followed by the thermodynamical universality of pure Lovelock black
hole \cite{dpp}, this is yet another new interesting universal
feature we have added. The Lovelock gravity is always trivial in the
critical dimension $d=2n+1$ with the corresponding curvature
$R^{(n)}_{abcd}$ vanishing, however it always represents a global
monopole for the Einstein gravity. What it means in general is that
the Lovelock degree $n$ does not matter for gravity in the critical
dimension. In the critical dimension gravity is always universal
including the BTZ black hole as well as its global monopole
description in the Einstein sector. This is indeed a very remarkable
general result.

We thank David Kastor for sharing his ideas.

\bibliographystyle{elsarticle-num}

\begin{thebibliography}{90}


\bibitem{bianchi} N. Dadhich, Pramana {\bf74}, 875 (2010) (arXiv:0802.3034)

\bibitem{dadhich1} N. Dadhich, {\it On "minimally curved spacetimes" in general relativity} (gr-qc/9705026); {\it On the Schwarzschild field} (gr-qc/9704068)

\bibitem{bv} M. Barriola and A. Vilenkin, Phys. Rev. Lett. {\bf63}, 341 (1989).

\bibitem{dny} N. Dadhich, K. Narayan and U. Yajnik, Pramana {\bf50}, 307 (1998) (gr-qc/9703034)

\bibitem{vac} N. Dadhich, Math Today {\bf 26}, 37 (2011) (arXiv:1006.0337); R. Cai and N. Ohta, Phys. Rev. D {\bf 74}, 064001 (2006) (hep-th/0604088);
R. Cai, L-Ming Cao, Y. Hu and S. Kim, Phys. Rev. {\bf 78}, 124012 (2008) (arXiv:0810.2610)

\bibitem{kastor} D. Kastor, {\it Th Reimann-Lovelock curvature tensor} (arXiv:1202.5287).

\bibitem{btz} M. Banados, C. Teitelboim and J. Zanelli, Phys. Rev. Lett. {\bf 69}, 1849 (1992).

\bibitem{dmk} N. Dadhich, A. Molina and A. Khugaev, Phys. Rev. {\bf 81}, 104026 (2010), (arXiv:1001.3922).

\bibitem{dpp} N. Dadhich, J. M. Pons and K. Prabhu, {\it Thermodynamical Universality of the Lovelock black hole}, (arXiv:1110.0673: {\it On the
static Lovelock black holes}, (arXiv:1201.4994).

\end{thebibliography}

\end{document}